\documentclass[aip,jap,reprint,graphicx]{revtex4-1}

\usepackage{graphicx}
\usepackage{amsmath}
\usepackage{textcomp} 
\usepackage{xcolor}
\usepackage{ulem}
\usepackage{amssymb}

\newcommand{\YIG}{Y$_\mathrm{3}$Fe$_\mathrm{5}$O$_\mathrm{12}$}

\begin{document}

\title{Spin waves in micro-structured yttrium iron garnet nanometer-thick films}

\author{Matthias~B.~Jungfleisch}
\email{jungfleisch@anl.gov}
\affiliation{Materials Science Division, Argonne National Laboratory, Argonne, Illinois 60439, USA}

\author{Wei~Zhang}
\affiliation{Materials Science Division, Argonne National Laboratory, Argonne, Illinois 60439, USA}

\author{Wanjun~Jiang}
\affiliation{Materials Science Division, Argonne National Laboratory, Argonne, Illinois 60439, USA}

\author{Houchen~Chang}
\affiliation{Department of Physics, Colorado State University, Fort Collins, Colorado 80523, USA}

\author{Joseph~Sklenar}
\affiliation{Department of Physics and Astronomy, Northwestern University, Evanston, Illinois 60208, USA}

\author{Stephen~M.~Wu}
\affiliation{Materials Science Division, Argonne National Laboratory, Argonne, Illinois 60439, USA}

\author{John~E.~Pearson}
\affiliation{Materials Science Division, Argonne National Laboratory, Argonne, Illinois 60439, USA}

\author{Anand~Bhattacharya}
\affiliation{Materials Science Division, Argonne National Laboratory, Argonne, Illinois 60439, USA}

\author{John~B.~Ketterson}
\affiliation{Department of Physics and Astronomy, Northwestern University, Evanston, Illinois 60208, USA}

\author{Mingzhong~Wu}
\affiliation{Department of Physics, Colorado State University, Fort Collins, Colorado 80523, USA}

\author{Axel~Hoffmann}
\affiliation{Materials Science Division, Argonne National Laboratory, Argonne, Illinois 60439, USA}

\date{\today}

\begin{abstract}

We investigated the spin-wave propagation in a micro-structured yttrium iron garnet wave\-guide of $40$~nm thickness. Utilizing spatially-resolved Brillouin light scattering microscopy, an exponential decay of the spin-wave amplitude of $(10.06 \pm 0.83)$~$\mu$m was observed. This leads to an estimated Gilbert damping constant of $\alpha=(8.79\pm 0.73)\times 10^{-4}$, which is larger than damping values obtained through ferromagnetic resonance measurements in unstructured films. The theoretically calculated spatial interference of waveguide modes was compared to the spin-wave pattern observed experimentally by means of Brillouin light scattering spectroscopy.
\end{abstract}

\maketitle
\section{Introduction}
\label{sec:intro}

Magnonics is an emerging field of magnetism studying the spin dynamics in micro- and nanostructured devices aiming for the development of new spintronics applications.\cite{Neusser,Magnonics,Magnonics_book} Up to now, ferromagnetic metals (for example, Permalloy and Heusler alloys) have been widely used for the investigation of magnetization dynamics on the nanoscale.\cite{Vogt_corner,Vogt_multiplexer,Demidov_JPD,Demido_open_end, Demidov_interference,Demidov_variable, Sebastian_Heusler} However, the Gilbert damping of Permalloy is two orders of magnitude higher than that of ferrimagnetic insulator yttrium iron garnet (YIG, \YIG). Recent progress in the growth of YIG films allows for the fabrication of low-damping nanometer-thick YIG films,\cite{Liu, Kelly, Pirro, Wu_YIG2012} which are well-suited for patterning of micro-structured YIG devices. This enables investigations of spin-wave propagation in plain YIG microstructures of sub-100~nm thicknesses which are a step forward for future insulator-based magnonics applications. 

In this work, we experimentally demonstrate spin-wave propagation in a micro-structured YIG waveguide of $40$~nm thickness and $4$~$\mu$m width. By utilizing spatially-resolved Brillouin light scattering (BLS) microscopy\cite{Vogt_corner, Vogt_multiplexer, Demidov_JPD, Sebastian_Heusler, Pirro} the exponential decay length of spin waves is determined. The corresponding damping parameter of the micro-structured YIG is estimated and compared to that determined from ferromagnetic resonance (FMR) measurements. Furthermore, we show that different spin-wave modes quantized in the direction perpendicular to the waveguide lead to a spatial interference pattern. We compare the experimental results to the theoretically expected spatial interference of the waveguide modes.

\section{Experiment}
\label{sec:exp}

\begin{figure}[b]
\includegraphics[width=0.85\columnwidth]{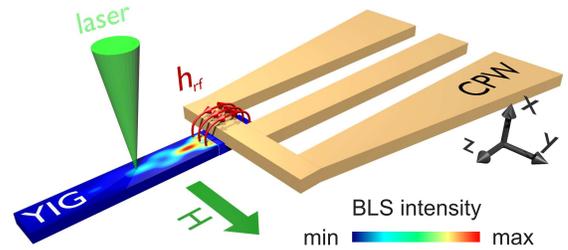}
\caption{\label{fig1} (Color online) Schematic illustration of the sample layout. The $4~\mu$m wide yttrium iron garnet waveguide is magnetized transversally by the bias magnetic field ${H}$. Spin waves are excited by a shortened coplanar waveguide and the spin-wave intensity is detected by means of spatially-resolved Brillouin light scattering microscopy. Colorbar indicates spin-wave intensity.} 
\end{figure}

\begin{figure}[t]
\includegraphics[width=0.72\columnwidth]{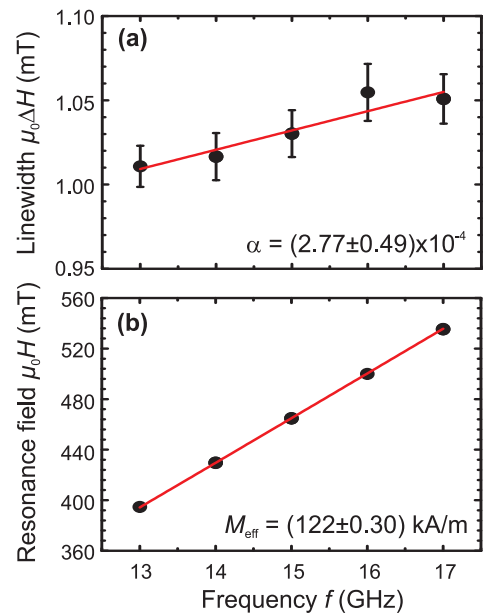}
\caption{\label{fig2} (Color online) (a) Ferromagnetic resonance peak-to-peak linewidth $\mu_0 \Delta H$ as a function of the resonance frequency $f$ of the unstructured $40$~nm YIG film. The red solid line represents a fit to Eq.~(\ref{FMR-alpha}). A Gilbert damping parameter $\alpha = (2.77\pm0.49)\times 10^{-4}$ is determined. (b) Ferromagnetic resonance field $\mu_0 H$ as a function of $f$. Error bars are smaller than the data symbols. } 
\end{figure}

Figure~\ref{fig1} shows a schematic illustration of the sample layout. The YIG film of $40$~nm thickness was deposited by magnetron sputtering on single crystal polished gadolinium gallium garnet (GGG, Gd$_3$Ga$_4$O$_{12}$) substrates of $500~\mu$m thickness with (111) orientation under high-purity argon atmosphere. The film was subsequently annealed in-situ at $800^\circ$C for 4~hours under an oxygen atmosphere of 1.12~Torr. The magnetic properties of the unstructured film were characterized by FMR: The peak-to-peak linewidth $\mu_0 \Delta H$ as a function of the excitation frequency $f$ is depicted in Fig.~\ref{fig2}(a). The Gilbert damping parameter $\alpha_\mathrm{FMR}$ can be obtained from FMR measurements using\cite{FMR}

\begin{equation}
\label{FMR-alpha}
\sqrt{3}\mu_0 \Delta H= \frac{2\alpha_\mathrm{FMR}}{\vert\gamma\vert} f + \mu_0 \Delta H_0,
\end{equation}	
where $\mu_\mathrm{0}$ is the vacuum permeability, $\gamma$ is the gyromagnetic ratio, $f$ is the resonance frequency and $\mu_0 \Delta H_0$ is the inhomogeneous linewidth broadening. We find a damping parameter of $\alpha_\mathrm{FMR} = (2.77\pm0.49)\times 10^{-4}$ [fit shown as a red solid line in Fig.~\ref{fig2}(a)]. The resonance field, $\mu_0 H$, as a function of the excitation frequency $f$ is shown in Fig.~\ref{fig2}(b). A fit to\cite{Azevedo_two_magnon}
\begin{equation}
\label{Kittel}
{f= \frac{\mu_\mathrm{0} \lvert\gamma\lvert}{2\pi} \sqrt{H_\mathrm{}(H_\mathrm{}+M_\mathrm{eff})}}
\end{equation}
yields an effective magnetization $M_\mathrm{eff}=(122\pm0.30)$~kA/m [solid line in Fig.~\ref{fig2}(b)]. In a subsequent fabrication process, YIG waveguides of $4~\mu$m width were patterned by photo-lithography and ion milling with an Ar plasma at 600~V for 5~min. In a last step, a shortened coplanar waveguide (CPW) made of Ti/Au (3~nm/150~nm) is patterned on top of the {YIG} waveguide (see Fig.~\ref{fig1}). The shortened end of the CPW has a width of 5~$\mu$m. The Oersted field of an alternating microwave signal applied to the CPW exerts a torque on the magnetic moments in the YIG and forces them to precess. The bias magnetic field is applied perpendicular to the short axis of the waveguide (Fig.~\ref{fig1}) providing efficient excitation of Damon-Eshbach spin waves. The microwave power $P_\mathrm{MW}= 1$~mW is sufficiently small to avoid possible perturbations of spin-wave propagation caused by nonlinearities.

\section{Discussions}
\label{sec:discussion}

\begin{figure}[t]
\includegraphics[width=1\columnwidth]{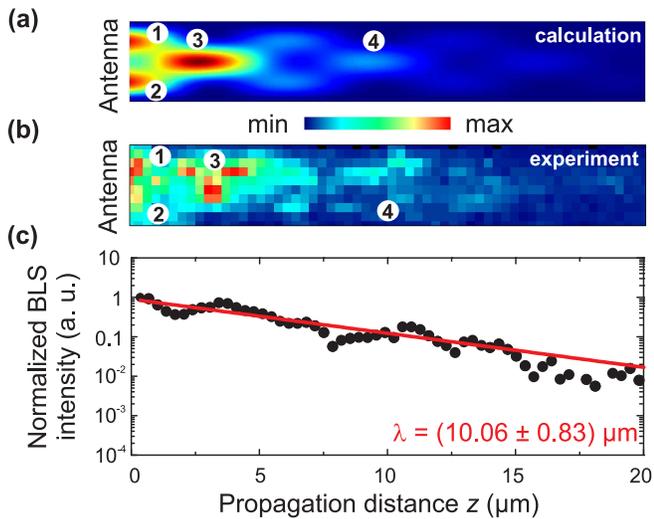}
\caption{\label{fig3} (Color online) (a) Calculated spatial interference pattern of the first two odd waveguide modes ($n=1$ and $n=3$. (b) Spatially-resolved BLS intensity map at an excitation frequency $f = 4.19$~GHz, applied microwave power $P_\mathrm{MW} = 1$~mW, biasing magnetic field $\mu_0 \Delta H = 83$~mT. {The numbers 1 -- 4 highlight the main features of the interference pattern.} (c) Corresponding BLS intensity integrated over the entire width of the YIG waveguide. An exponential decay of the spin-wave amplitude $\lambda = (10.06 \pm 0.83)~\mu$m is found. } 
\end{figure}

\begin{figure}[b]
\includegraphics[width=0.91\columnwidth]{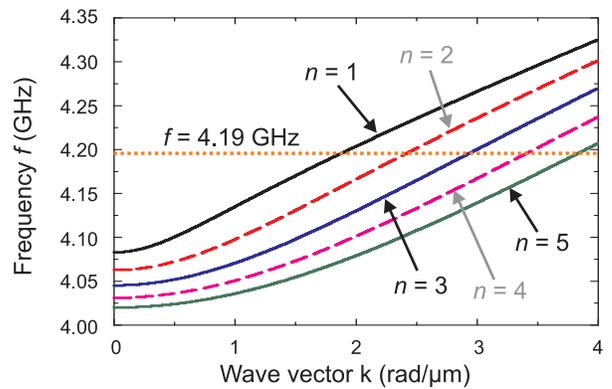}
\caption{\label{fig4} (Color online) Dispersion relations the first five waveguide modes of a transversally magnetized YIG stripe.\cite{parameters} Only modes with a odd quantization number $n$ can be excited (solid lines). $f$ denotes the excitation frequency.}
 \end{figure}

In order to detect spin-wave propagation in the YIG waveguide spatially-resolved BLS microscopy with a resolution of $250$~nm is employed. 
To characterize the propagating spin waves, the BLS intensity was recorded at different distances from the antenna. A spatially-resolved BLS intensity map is shown in Fig.~\ref{fig3}(b) at an exemplary excitation frequency of $f=4.19$~GHz. Spin waves are excited near the antenna and propagate towards the opposite end of the waveguide. To further analyze the data and to minimize the influence of multi-mode propagation in the YIG stripe (this will be discussed below), the BLS intensity is integrated over the width of the waveguide. The corresponding BLS intensity as a function of the distance from the antenna is illustrated in Fig.~\ref{fig3}(c). The decay of the spin-wave amplitude can be described by: \cite{Sebastian_Heusler,Pirro}

\begin{equation}
\label{exp_decay}
I(z)=I_0 \mathrm{e}^{-\frac{2z}{\lambda}}+ b,
\end{equation}	
where $z$ is the distance from the antenna, $\lambda$ is the decay length of the spin-wave amplitude and $b$ is an offset. From Fig.~\ref{fig3}(b), it is apparent that the data-points follow an exponential behavior. A fit according to Eq.~(\ref{exp_decay}) yields the decay length $\lambda$. For an excitation frequency of $f = 4.19$~GHz we find $\lambda = (10.06\pm 0.83)~\mu$m. This value is larger than decay lengths reported for Permalloy (\textless$6~\mu$m, see Ref.~\onlinecite{Demidov_interference,Pirro_solidi,Akerman}), but it is smaller than the largest decay length found for the Heusler-compound Co$_2$Mn$_{0.6}$Fe$_{0.4}$Si ($8.7$ -- $16.7~\mu$m, see Ref.~\onlinecite{Sebastian_Heusler}). Pirro et al. reported a decay length of $31~\mu$m in thicker YIG waveguides (100~nm) grown by liquid phase epitaxy (LPE) with a $9$~nm thick Pt capping layer.\cite{Pirro} In order to understand this discrepancy between our and their results, one has to take into account two facts: (1) State-of-the-art LPE fabrication technology can not be employed to grow film thicknesses below $\sim100$~nm. To date, sputtering offers an alternative approach to grow sub-100~nm thick YIG films with a sufficient quality.\cite{Houchen_IEEE} (2) Taking into account the spin-wave group velocity $v_\mathrm{g}=\partial \omega /\partial k$ and the spin-wave lifetime $\tau$, the theoretically expected decay length $\lambda$ can be calculated from $\lambda = v_\mathrm{g}\cdot\tau$. The group velocity $v_\mathrm{g}$ can be derived directly from the dispersion relation (see Fig.~\ref{fig4}). A thinner YIG film has flatter dispersion and a smaller $v_\mathrm{g}$. Consequently, the expected decay length is smaller for thinner YIG samples and so it is natural that the decay length reported here is shorter than the one found in Ref.~\onlinecite{Pirro} for 100~nm thick YIG waveguides.

We estimate the group velocity from the spin-wave dispersion to be $v_\mathrm{g} = 0.35-0.40~\mu$m/ns. Using our experimentally found decay length, the spin-wave lifetime is determined to be $\tau = 29$~ns. We can use the BLS-data to determine the corresponding Gilbert damping parameter $\alpha_\mathrm{BLS}$. In case of Damon-Eshbach spin waves, the damping is given by\cite{Stancil}

\begin{equation}
\label{alpha}
\alpha_\mathrm{BLS}= \frac{1}{\tau} (\frac{\gamma \mu_0 M_\mathrm{eff}}{2} + 2\pi f)^{-1}.
\end{equation}	

A Gilbert damping parameter of the micro-structured YIG waveguide obtained by BLS characterization is found to be $\alpha_\mathrm{BLS} = (8.79\pm 0.73)\times 10^{-4}$, which is a factor of 3 times larger than that determined by FMR in the unstructured film [$\alpha_\mathrm{FMR}= (2.77\pm 0.49)\times 10^{-4}$]. This difference might be attributed to the micro-structuring of the YIG waveguide by Ar ion beam etching. The etching might enhance the roughness of the edges of the YIG waveguides and the resist processing could have an influence on the surface quality\cite{McMorran,Roshchupkina} which could possibly lead to an enhancement of the two-magnon scattering process.\cite{Arias} {It would be desirable to perform FMR measurements on the YIG waveguide. However, since the structured bar is very small, the FMR signal is vanishingly small which makes it difficult to determine the Gilbert damping in this way.}

While the discussion above only considered the BLS-intensity integrated over the waveguide width, we will focus now on the spatial interference pattern shown in Fig.~\ref{fig3}(b). The spin-wave intensity map can be understood by taking into account the dispersion relation of magnetostatic spin waves in an in-plane magnetized ferromagnetic thin film (see Fig.~\ref{fig4}). Due to the lateral confinement, the wave vector is quantized across the width of the YIG waveguide, $k_y=n\pi/w$, where $n \in\mathbb{N}$. The wave vector $k_z$ along the long axis of the waveguide ($z$-axis) is assumed to be non-quantized. We follow the approach presented in Ref.~\onlinecite{Demidov_interference}. The dynamic magnetization is assumed to be pinned at the edges of the waveguide which can be considered by introducing an effective width of the waveguide. \cite{Demidov_interference,Guslienko} 

Figure~\ref{fig4} shows the calculated dispersion relations of different spin-wave modes quantized across the width of the strip. The dashed line represents a fixed excitation frequency. At a particular frequency different spin-waves modes with different wave vectors $k_z$ are excited simultaneously. This leads to the occurrence of spatially periodic interference patterns. In the present excitation configuration, only modes with an odd quantization number $n$ can be excited ($n$ determines the number of maxima across the width of the waveguide). \cite{} Since the intensity of the dynamic magnetization of these modes decreases with increasing $n$ as $1/n^2$, we only consider the first two odd modes $n=1$ and $n=3$.

According to Ref.~\onlinecite{Demidov_interference} the spatial distribution of the dynamic magnetization of the $n$-th mode can be expressed as
\begin{equation}
\label{interference_1}
m_{n}(y,z)\propto \mathrm{sin}(\frac{n\pi}{w}y)\mathrm{cos}(k_{z}^{n}z-2\pi f t +\phi_{n}),
\end{equation}
where $f$ is the excitation frequency, $k_{z}^{n}$ is the longitudinal wave vector of the $n$-th spin-wave mode and $\phi_{n}$ is the phase. \cite{Schneider_phase} The spin-wave intensity distribution $I_n$ of the $n$-th mode can be derived by averaging $m_n (y,z)^2$ over one oscillation period $1/f$. The entire interference pattern can be obtained from the same procedure using the sum $m_1 (y,z) +\frac{1}{3} m_3 (y,z)$. The factor 1/3 accounts for the lower excitation efficiency of the $n=3$ mode. Thus, the intensity is given by \cite{Demidov_interference}
\begin{equation}
\label{interference_2}
\begin{split}
I_\Sigma (y,z) \propto \mathrm{sin}(\frac{\pi}{w}y)^2+\frac{1}{9} \mathrm{sin}(\frac{3\pi}{w}y)^2 \\+ \frac{2}{3} \mathrm{sin}(\frac{\pi}{w}y)\mathrm{sin}(\frac{3\pi}{w}y)\mathrm{cos}&(\Delta k_{z}z+ \Delta\phi),
\end{split}
\end{equation}
where $\Delta k_z = k_z^3-k_z^1$ and $\Delta \phi = \phi_3 - \phi_1$. This pattern repeats periodically. The phase shift $\phi$ shifts the entire pattern along the $z$-direction and the wave-vector difference $\Delta k_z= 0.97$~rad/$\mu$m can be calculated from the dispersion relation (Fig.~\ref{fig4}). The calculated spatial interference pattern is depicted in Fig.~\ref{fig3}(a) using $\Delta \phi =0$ and taking into account for the exponential decay of the spin-wave amplitude by multiplying Eq.~(\ref{interference_2}) with $e^{-2z/\lambda_\mathrm{}}$ using the experimentally determined $\lambda= 10.03~\mu$m. As is apparent from Fig.~\ref{fig3}(a) and (b) a qualitative agreement between calculation and experiment is found. {(The numbers 1 -- 4 highlight the main features of the interference pattern in experiment and calculation.) The small difference in  Fig.~\ref{fig3}(a) and (b) can be explained by considering the fact that in the calculation a spin-wave propagation at an angle of exactly 90$^\circ$ (Damon-Eshbach configuration) with respect to the antenna/external magnetic field is assumed. However, in experiment small misalignments of the external magnetic field might lead to a small asymmetry in the interference pattern.}

\section{Conclusion}
\label{sec:conclusion}
 
In summary, we demonstrated spin-wave excitation and propagation in micro-fabricated pure YIG wave\-guides of 40~nm thickness. BLS-characterization revealed a decay length of the spin-wave amplitude of 10~$\mu$m leading to an estimated Gilbert damping parameter of $\alpha_\mathrm{BLS} = (8.79\pm 0.73)\times 10^{-4}$. This value is a factor 3 larger than the one determined for the unstructured YIG film by means of FMR techniques [$\alpha_\mathrm{FMR}=(2.77\pm0.49)\times 10^{-4}$]. The difference might be attributed to micro-structuring using ion beam etching. The observed spatial spin-wave intensity distribution is explained by the simultaneous excitation of the first two odd waveguide modes. These findings are important for the development of new nanometer-thick magnon spintronics applications and devices based on magnetic insulators.



%
%
%
%
%
%
%
%
%

\section{Acknowledgments}
\label{sec:ack}

Work at Argonne was supported by the U.S. Department of Energy, Office of Science, Materials Science and Engineering Division. Work at Colorado State University was supported by the U.S. Army Research Office, and the U.S. National Science Foundation. Lithography was carried out at the Center for Nanoscale Materials, which is supported by DOE, Office of Science, Basic Energy Sciences under Contract No. DE-AC02-06CH11357.


\begin{thebibliography}{19}
\bibitem{parameters} For the calculation of the dispersion relations the following parameters have been used: external magnetic field $\mu_0 H = 83$~mT, exchange constant $A= 3.6$~pJ/m, saturation magnetization $M_\mathrm{S}$=140~kA/m, effective width of the waveguide $w_\mathrm{eff}=3.5~\mu$m, YIG-film thickness $t = 40$~nm.
\end{thebibliography}
\end{document}